\documentclass[runningheads]{llncs}
\usepackage{graphicx}
\usepackage{placeins}
\usepackage[T1]{fontenc}
\usepackage{listings}
\usepackage{balance}
\usepackage{booktabs}
\usepackage{todonotes}
\usepackage{graphicx}
\usepackage{amsmath}
\usepackage{subcaption}
\usepackage{pdflscape}
\usepackage{adjustbox}
\usepackage{geometry}
\usepackage{xcolor}
\usepackage{hyperref}

\hypersetup{
  colorlinks=true,
  citecolor=black,
  linkcolor=green,
  urlcolor=blue,
  pdfborder={0 0 1},
  pdfborderstyle={/S/U/W 1}
}

\let\oldcite\cite
\renewcommand{\cite}[1]{\fcolorbox{green}{white}{\oldcite{#1}}}

\let\oldref\ref
\renewcommand{\ref}[1]{\fcolorbox{green}{white}{\oldref{#1}}}

\begin{document}
\title{A Short Survey on Formalising Software Requirements with Large Language Models\thanks{Supported by ADAPT research centre, Ireland}}
%
%\titlerunning{Abbreviated paper title}
% If the paper title is too long for the running head, you can set
% an abbreviated paper title here
%
\author{Arshad Beg\inst{1}\orcidID{0009-0004-6939-0411} \and 
Diarmuid O'Donoghue\inst{1}\orcidID{0000-0002-3680-4217} \and 
Rosemary Monahan\inst{1}\orcidID{0000-0003-3886-4675}}

\institute{Department of Computer Science, Maynooth University, Ireland \\
\email{arshad.beg@mu.ie, diarmuid.odonoghue@mu.ie, rosemary.monahan@mu.ie}
}
\maketitle              % typeset the header of the contribution

\begin{abstract}

This paper presents a focused literature survey on the use of large language models (LLM) to assist in writing formal specifications for software. A summary of thirty-five key papers is presented, including examples for specifying programs written in Dafny, C and Java. This paper arose from the project “VERIFAI: Traceability and verification of natural language requirements” that addresses the challenges in writing formal specifications from requirements that are expressed in natural language. Our methodology employed multiple academic databases to identify relevant research. The AI-assisted tool Elicit facilitated the initial paper selection, which were manually screened for final selection. The survey provides valuable insights and future directions for utilising LLMs while formalising software requirements.

\end{abstract}

\section{Introduction}

Relying on natural language in software requirements  often leads to ambiguity. In addition, requirements which are not expressed in a formal mathematical notation cannot be  guaranteed through formal verification techniques as required to meet standards e.g. \cite{10.1145/2070336.2070341,6136916,10.5555/1151816.1151817} in safety critical software. Expressing requirements in formal notation requires training in the domain of requirements engineering, as well as knowledge of formal notation and associated proof methods, often increasing the software development cycle time by a factor of 30\% \cite{Huisman2024}. In this project, we aim to ease the burden of writing specification, helping to bridge the gap between the need for formal verification techniques and their lack of use in the software industry (due to the fast pace of the industry environment). 

Formalising software requirements ensures clarity, correctness and verifiability, and requires formal specification languages, logic and verification techniques, such as theorem proving and model-checking, to guarantee the correctness of the software system under construction. Like all other fields, the development of Large Language Models (LLMs) has opened a world of opportunities where we can exploit their power to generate formal requirements and accompanying specifications. 

In this paper, we present the results of our structured literature review which examines how large language models are currently used to assist in writing formal specifications. Main research questions for conducting systematic literature review on the topic are as follows: \\
\textbf{RQ1:} What methodologies leverage Large Language Models (LLMs) to transform natural language software requirements into formal notations? \\
% while addressing the associated challenges and limitations identified by researchers? - [addressed in section \ref{literature_formalisation_llms}]\\
\textbf{RQ2:} What are the emerging trends and future research directions in using LLMs for software requirements formalisation? 
%- [addressed in section \ref{sec:future_directions} with discussion in succeeding sub-section \ref{subsec:CoTnPromptEngg}]

We represent a motivating example of generating formal requirements and specifications in section \ref{sec:example}. Section \ref{sec:methodologyforsurvey} presents our methodology for the literature review. A summary of contribution from each paper addressing \textbf{RQ1} is provided in section \ref{literature_formalisation_llms}. Section \ref{sec:future_directions} presents future directions based on our findings and the discussion on chain of thought and prompt engineering in sub-section \ref{subsec:CoTnPromptEngg}, addressing \textbf{RQ2}. Concluding remarks are provided in Section \ref{sec:conclusions}.

\section{Example of Generating Assertions}
\label{sec:example}

Dafny is a verification-aware programming language that has native support for expressing formal specifications and is equipped with a static program verifier to automatically verify implementations against specifications using deductive verification.  Mugnier et al. \cite{mugnier2024laurelgeneratingdafnyassertions} present a framework named Laurel to generate Dafny assertions using LLMs, using two domain-specific prompting techniques. The first technique locates the position in code where an assertion, which provides part of the formal specifications, is missing. This is done through analysis of the Dafny verifier's error message. A placeholder is inserted  at the particular location where the assertion is missing. The second technique involves the provision of example assertions from a codebase. Here, Laurel is able to generate over 50\% of the required helper assertions, making it a viable approach to deploy, for automating the program specification and verification process. 

\subsection{Generating Dafny Assertions}

We illustrate an example using a Dafny lemma as reported in \cite{mugnier2024laurelgeneratingdafnyassertions}, where a helper assertion is needed in the verification process. Lemmas are used in Dafny to specify properties that may be used in the verification process. The lemma, in this example (the full body of Lemma is included in Appendix \ref{dafny_lemma}, ensures that the integer and fractional parts are correctly extracted while parsing a decimal string.
 
\noindent{Explaining the lemma presented, we have:}

\noindent\textbf{Precondition} (\textbf{requires} clause): The function assumes that every character in \texttt{s1} is a digit ('0' to '9').  

\noindent\textbf{Postcondition} (\textbf{ensures} clause): The function guarantees that when \texttt{ParseDecStr} is applied to the concatenation of \texttt{s1} and \texttt{s2}, the first part of the result remains \texttt{s1}, and the second part contains \texttt{"." + s2}.  

\noindent\textbf{Base Case}: If \texttt{s1} contains a single character, the function asserts that parsing \texttt{"."+s2} leads to an empty integer part.  

\noindent\textbf{Recursive Case}: The function calls itself with the tail of \texttt{s1} to process it recursively. However, to help the verifier understand the transformation, an assertion is added:  \begin{quote}  
\texttt{\textbf{assert} s1 + "." + s2 == [s1[0]] + (s1[1..] + "." + s2);}
  \end{quote}
This assertion explicitly states how the string is decomposed and aids the SMT solver in proving correctness.

\noindent\textbf{Role of the Helper Assertion:}

Without the assertion, the Dafny verifier struggles to establish the correctness of the postcondition due to the complexity of reasoning about string concatenation. The assertion serves as an intermediate step, breaking down the transformation into a form that is easier for the solver to handle.

This example motivates our work by demonstrating the importance of inserting specification in the form of helper assertions in Dafny proofs. Tools like Laurel \cite{mugnier2024laurelgeneratingdafnyassertions} aim to automate this process by leveraging Large Language Models to determine the relevant assertions.

\section{Methodology for Literature Review}
\label{sec:methodologyforsurvey}
To conduct a structured and thorough review of literature on Natural Language Processing (NLP), Large Language Models (LLMs), and their use in software requirements, the following approach is followed. Several academic databases, including IEEE Xplore, ACM Digital Library, Scopus, Springer Link, and Google Scholar, are searched using specific keywords. The core search terms include “NLP,” “LLMs,” and “Software Requirements,” with broader terms such as “specification,” “logic,” “verification,” “model checking,” and “theorem proving” used to expand the scope. The number of results differs notably across databases. For example, IEEE Xplore returns 17 peer-reviewed articles, Scopus lists 20, Springer Link filters to 595, ACM Digital Library provides 1,368 results, and Google Scholar shows 14,800 references since 2021. These discrepancies highlight the importance of applying precise selection methods to extract the most relevant studies.

%\dodnote{Thinking as a review: they will ask who says Elicit is a good tool for this? Why is it better than the previous approach? Do we need to write defensively here? Can one of our results be an evaluation of Elicit? (ratio of relevant papers)}

 %\rmnote{what was the criteria used for inclusion, exclusion etc.}

To streamline the process of locating strong contributions, the AI-powered tool Elicit \cite{elicit_tool} is used. Elicit supports the literature review by offering summarised content and DOIs for suggested papers. While it helps reduce the manual workload during the initial phase, every suggested paper in this review is manually reviewed to ensure its relevance. This ensures that the final list excludes any unrelated or off-topic material. After the initial filtering, a manual review is performed to confirm the relevance and quality of each paper. Abstracts are first assessed to judge suitability. If the abstract lacks clarity or depth, a further examination of the full text is conducted.
%Elicit has shown a good level of accuracy in delivering relevant research results, making the search process smooth and generally reliable. In our experience, every time a provided DOI was accessed, it led to a legitimate, peer-reviewed research paper, which reflects well on the tool’s consistency. 
% While it's important to remain cautious with any AI tool, Elicit has so far provided a dependable experience, especially for those looking to explore academic content with a balance of convenience and credibility.
%Although Elicit assists with the early stages of reference gathering, the final decisions remain in the hands of the researchers. For each section of this survey, Elicit is employed only for identifying references, maintaining a clear balance between automation and expert judgement.

%After the selection, a detailed manual evaluation is carried out to confirm that each chosen paper provides substantial input to the field. 
Abstracts are closely read, and when necessary, the full text is reviewed using the following exclusion and inclusion criteria:

%This ensures that only papers offering significant value to the discussion on NLP, LLMs, and software requirements are included. From the larger collection of retrieved literature, selected works %are shortlisted based on their contribution, citation count, and alignment with the aims of this review. These are presented in Section \ref{literature_formalisation_llms}. Manual checks guarantee %that all included articles offer strong theoretical or empirical foundations, and that no redundant or weak contributions are carried over.
%\abnote{Inclusion and Exclusion Criteria}
%Based on above methodology, we sketch exclusion and inclusion criteria explicitly as follows:

\textbf{Inclusion Criteria:} Studies are included if they offer meaningful theoretical or empirical insights related to NLP, LLMs, and their application in software requirements. This includes topics like specification, formal logic, verification, and formal methods. 

\textbf{Exclusion Criteria:} Papers are excluded if they show in-sufficient relevance to the intersection of NLP/LLMs and software requirements, or if their abstracts or full texts lack sufficient detail. Non-peer-reviewed materials, duplicates, and items suggested by Elicit but deemed irrelevant after manual review are also removed.
%The methodology balanced automation with human judgment, ensuring that the final set of papers captured critical advancements in the field. The survey not only relies on published peer-reviewed research but also follows an iterative approach where insights from the reviewed literature guide further keyword adjustments and additional searches. This iterative process enhances the comprehensiveness of the survey, ensuring a well-rounded perspective on the state-of-the-art in NLP, LLMs, and software requirements.

\section{Formalising Requirements through LLMs}
\label{literature_formalisation_llms}

\begin{figure}[htb]
    \centering
    \includegraphics[width=\textwidth, keepaspectratio]{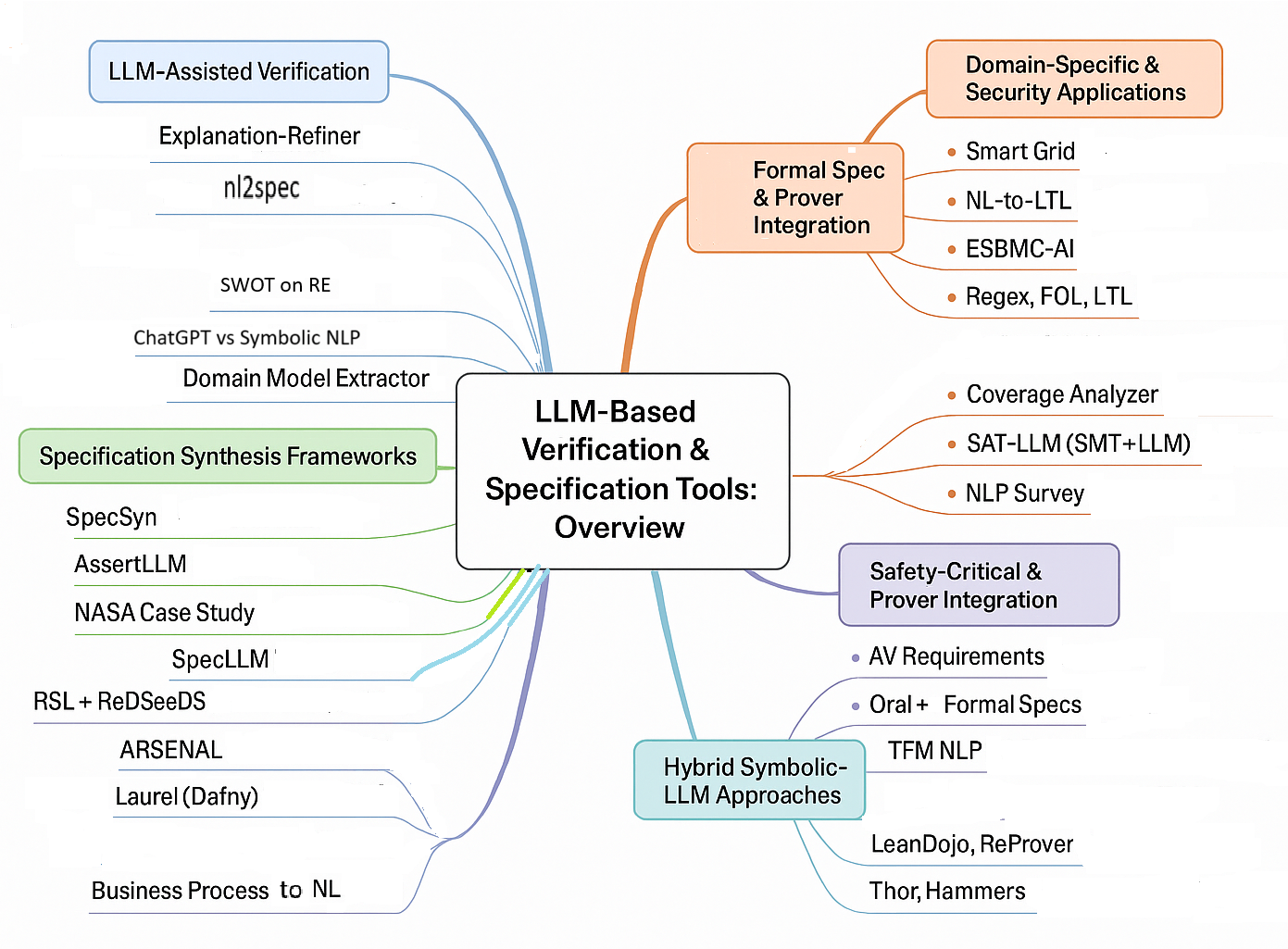}
    \caption{LLM-Based Verification and Specification Literature Overview}
		\label{Fig:LitClassification}
\end{figure}

This section provides an overview of the literature surveyed. A summary table is included in Appendix \ref{appendix:summarytables}, listing each citation along with the tool, technique, or framework developed or analysed in the paper, accompanied by a brief description of its contribution. Figure~\ref{Fig:LitClassification} presents a tree-structured overview of the studied literature.

Table \ref{Tab:LitClassification} presents a classification of the surveyed literature based on their methodological approach. Most works fall under the ``Prompt-only'' category, where LLMs are used without further tuning. Some studies involve human-in-the-loop or iterative prompting, while others fine-tune LLMs for improved performance. A set of works integrates LLMs with verifiers, and a few adopt neuro-symbolic methods combining LLMs with SMT solvers or theorem provers. Other categories include baseline/manual approaches using controlled natural language, meta-analyses and tool support studies, and a single work proposing IDE integration. Table \ref{Tab:LitClassification} is a condensed version of Table \ref{tab:classification} found in the Appendix. To provide readers with a more comprehensive reference, we have also included Table \ref{tab:merged_classification} in the Appendix. This table integrates the developed tools, their methodological classifications, and a brief summary of each work, offering a clear overview of the purpose, approach, and contributions of each paper at a glance.  

\begin{table}[ht]
\caption{Classification of Surveyed Literature by Methodology}
\label{Tab:LitClassification}
\resizebox{\textwidth}{!}{%
\begin{tabular}{|p{6.0cm}|p{6.5cm}|}
\hline
\textbf{References} & \textbf{Classification} \\
\hline
\cite{10.1145/2070336.2070341}, \cite{6136916}, \cite{10500073}, \cite{quan2024verificationrefinementnaturallanguage}, \cite{arora2023advancingrequirementsengineeringgenerative}, \cite{reinpold2024exploringllmsverifyingtechnical}, \cite{10684640}, \cite{Granberry2025a}, \cite{Jason2022}, \cite{BoraCaglayan2024}, \cite{wu2024lemurintegratinglargelanguage}, \cite{DBLP:journals/corr/abs-2305-09993}, \cite{DBLP:conf/emnlp/ShumDZ23}, \cite{Granberry2025}, \cite{zhang2023selfconvincedpromptingfewshotquestion}, \cite{mandal2023largelanguagemodelsbased}, \cite{10500073}, \cite{quan2024verificationrefinementnaturallanguage}, \cite{10656469}, \cite{zhang2023selfconvincedpromptingfewshotquestion}, \cite{Jason2022} & Prompt-only \\
\hline
\cite{quan2024verificationrefinementnaturallanguage}, \cite{10500073}, \cite{10684640}, \cite{DBLP:conf/emnlp/ShumDZ23}, \cite{BoraCaglayan2024} & Prompt + Iterative / Human-in-loop / CoT \\
\hline
\cite{10.5555/1151816.1151817}, \cite{nowakowski2013requirements}, \cite{7092662}, \cite{10207159}, \cite{Kojima2022}, \cite{10684640}, \cite{7092662} & Fine-tuned \\
\hline
\cite{cosler2023nl2specinteractivelytranslatingunstructured}, \cite{mandal2023largelanguagemodelsbased}, \cite{Ghosh2016}, \cite{tihanyi2024newerasoftwaresecurity}, \cite{wu2024lemurintegratinglargelanguage}, \cite{zhang2023selfconvincedpromptingfewshotquestion}, \cite{BoraCaglayan2024} & Verifier-in-loop \\
\hline
\cite{10.1002/spe.430}, \cite{Req2SpecPaper}, \cite{quan2024verificationrefinementnaturallanguage} & Neuro-symbolic (LLM + SMT/Theorem Prover) \\
\hline
\cite{li2024specllmexploringgenerationreview}, \cite{491451}, \cite{nowakowski2013requirements}, \cite{Huisman2024}, \cite{li2024specllmexploringgenerationreview}, \cite{491451} & Baseline / Manual / Controlled NL \\
\hline
\cite{10207159}, \cite{10.1145/2976767.2976769}, \cite{ernst:LIPIcs.SNAPL.2017.4}, \cite{Ma2024}, \cite{10.1145/2976767.2976769} & Meta-analysis / Tool Support \\
\hline
\cite{6823180} & IDE Integration Proposal \\
\hline
\end{tabular}%
}
\label{tab:classification_grouped}
\end{table}

\subsection{LLM-Assisted Verification and Specification Tools}
The paper \cite{10500073} proposes using LLMs, like GPT-3.5, to verify code by analysing requirements and explaining whether they are met. 
The work \cite{quan2024verificationrefinementnaturallanguage} provides verification and refinement of natural language explanations by making LLMs and theorem provers work together. A neuro-symbolic framework i.e. Explanation-Refiner is represented. LLMs and theorem provers are integrated together to formalise explanatory sentences. The theorem prover then provides the guarantee of validated sentence explanations. Theorem prover also provides feedback for further improvements in NLI (Natural Language Inference) model. Error correction mechanisms can also be deployed by using the tool Explanation-Refiner. Consequently, it automatically enhances the quality of explanations of variable complexity.  
The work \cite{fan2025evaluatingabilitylargelanguage} evaluated GPT-4o's ability to generate specifications for C programs that can be verified using VeriFast, a static verifier based on separation logic. Their experiments, which use different user inputs and prompting techniques, show that while GPT-4o's specifications maintain functional behaviour, they often fail verification and include redundancies when verifiable.

\subsection{Frameworks for Specification Synthesis}
The work \cite{cosler2023nl2specinteractivelytranslatingunstructured} details about nl2spec, a framework that leverages LLMs to generate formal specifications from natural language, addressing the challenge of ambiguity in system requirements. Users can iteratively refine translations, making formalization easier. The work \cite{cosler2023nl2specinteractivelytranslatingunstructured} provides an open-source implementation with a web-based interface. An automatic synthesis of software specifications is provided through LLMs in \cite{mandal2023largelanguagemodelsbased}. Work \cite{mandal2023largelanguagemodelsbased} proposed SpecSyn framework that uses an advanced language model to automatically generate software specifications from natural language text. It treats specification generation as a sequence-to-sequence learning task and outperforms previous tools by 21\% in accuracy, extracting specifications from both single and multiple sentences.

The paper \cite{Req2SpecPaper} introduced Req2Spec, an NLP-based tool that analyses natural language requirements to create formal specifications for \verb"HANFOR", a large-scale requirements and test generation tool. Tested on 222 automotive software requirements at BOSCH, it correctly formalized 71\% of them. The work \cite{Ma2024} represents a novel framework named SpecGen to generate specifications through LLMs. Two phases are applied. First phase is about having prompts in conversational style. Second phase is deployed where correct specifications are not generated. Here, four mutation operators are applied to ensure the correctness of the generated specifications. Two benchmarks i.e. SV-COMP and SpecGen are used. Verifiable specifications are generated successfully for 279 out of 384 programs, making \cite{Ma2024} a viable approach.

\subsection{Domain-Specific and Security-Focused Applications}
AssertLLM tool is presented in \cite{10691792}. The tool generates assertions to do hardware verification from design specifications, exploiting three customised LLMs. It is done in three phases, first understanding specifications, mapping signal definitions and generating assertions. The results show that AssertLLM produced 89\% correct assertions with accurate syntax and function. The work \cite{10.1002/spe.430} reports on formal verification of NASA's Node Control Software natural language specifications. The software is deployed at International Space Station. Errors found in the natural language requirements are reported by the authors with a commentary on lessons learnt.

SpecLLM \cite{li2024specllmexploringgenerationreview} explores the space of generating and reviewing VLSI design specifications with LLMs. The utility of LLMs is explored with the two stages i.e. (1) \textbf{generation} of architecture specifications from scratch and from register transfer logic (RTL) code; and (2) \textbf{reviewing} these generated specifications. In \cite{reinpold2024exploringllmsverifyingtechnical}, the potential and power of LLMs is exploited for smart grid requirement specifications improvement. Here, the performance of GPT-4o and Claude 3.5 Sonnet is analysed through f1-scores, achieving in range of 79\% - 94\%.

\subsection{Formal Specification Translation and Evaluation}
In \cite{10207159}, symbolic NLP and ChatGPT performance is compared while generating correct formal contracts written in the Java Modelling Language (JML) which have been extracted from natural language specifications. The paper \cite{10684640} reports the translation between NL and Linear Temporal Logic (LTL) formulas through the use of LLMs. Dynamic prompt generation and human interaction with LLMs are amalgamated to deal with the mentioned challenges. Unstructured natural language requirements are converted to NL-LTL pairs. The approach achieved up to 94.4\% accuracy on publicly available datasets with 36 and 255,000 NL-LTL pairs. The primitive work \cite{Nelken1996} described automatic translation from natural language sentences to temporal logic, in order to deploy formal verification of the requirements.

\subsection{Safety-Critical and Prover-Integrated Applications}
The work \cite{Yang2023} introduced LeanDojo, an open-source toolkit that enables programmatic interaction with Lean theorem prover. Using LeanDojo’s extracted data, \cite{Yang2023} developed ReProver, a retrieval-augmented LLM-based prover. Thor \cite{Jiang2022} is a framework developed to integrate language models with theorem provers. It improved the accuracy from 39\% to 57\% using the PISA dataset and also outperformed previous works on the MiniF2F dataset. The paper \cite{Granberry2025} explores integrating Co-pilot with formal methods. The integration of Copilot and formal methods is proposed through development of IDE containing language servers. Granberry et al. \cite{Granberry2025a} explored combining LLMs with symbolic analysis to generate specifications for C programs. They enhanced LLM prompts using outputs from PathCrawler and EVA to produce ACSL annotations.

\subsection{Hybrid Symbolic-LLM Methodologies and Future Outlook}
SAT-LLM, a unique framework to remove conflicting requirements is represented in \cite{10.1145/3691620.3695302}. It integrated Satisfiability Modulo Theories (SMT) solvers with LLMs. SAT-LLM performed better than ChatGPT alone, identifying 80\% of conflicts with a Precision of 1.00, Recall of 0.83, and an F1 score of 0.91. \cite{arora2023advancingrequirementsengineeringgenerative} outlines key research directions for the stages of software requirement engineering, conducts a SWOT analysis, and share findings from an initial evaluation. 

The purpose of Dafny is to automate proofs by outsourcing them to an SMT solver. \cite{mugnier2024laurelgeneratingdafnyassertions} presented a framework named Laurel to generate Dafny assertions using LLMs. Laurel was able to generate over 50\% of the required helper assertions. \cite{ernst:LIPIcs.SNAPL.2017.4} used input of error messages, variable names, procedure documentation and user questions. They discussed available literature for generating assertions by synthesising sentences in testing phase.

\subsection{Historical and Foundational Work}
The paper \cite{nowakowski2013requirements} introduced a model-based language (Requirements Specification Language - RSL). The framework functionality is integrated as development platform named ReDSeeDS. A similar work is reported in \cite{Ghosh2016} which describes the ARSENAL framework and methodology designed to perform automatic requirements specification extraction from natural language. An interesting work is presented in \cite{6823180}. It includes generation of natural language from business process models. The generated natural language is found complete and more understandable. In primitive work of 1996 \cite{491451}, software requirements were expressed in a limited set of natural language referred to as controlled natural language. The primitive work in the domain is about RML \cite{Greenspan1986}. RML bundled with features of writing requirements, which are based on conceptual model.

\subsection{Evaluation of the literature surveyed}

The literature surveyed in section \ref{literature_formalisation_llms} suggests that assertion generation currently shows higher reliability than full contract synthesis. Tools like Laurel and AssertLLM demonstrate success rates above 50\% and 89\%, respectively, when generating helper assertions or verifying hardware designs. These assertions typically focus on specific program points or behaviors, making them easier for LLMs to manage. In contrast, full contract generation, such as translating natural language to formal Java Modeling Language (JML) contracts or complete temporal logic formulas, often results in partially correct or unverifiable outputs. For example, efforts using GPT-4o to produce VeriFast-compatible specifications found that generated contracts often failed formal verification checks despite being functionally reasonable.

There is a clear trend where tasks involving smaller, well-scoped units like assertions yield more accurate results from LLMs. This is likely due to the limited context and reduced complexity compared to full contract generation. Full formal specifications, especially across multiple sentences or involving system-wide properties, require greater abstraction and contextual understanding. While frameworks like SpecGen and SpecSyn have made progress, their outputs often need iterative refinement or mutation operators to reach acceptable accuracy. This contrasts with tools like AssertLLM or Laurel, where results are immediately usable or require minimal correction.

For future outlook, this short paragraph is based on our discussion of Section \ref{sec:future_directions} of the paper as well. We can say that the combination of LLMs with formal methods (e.g., theorem provers or SMT solvers) shows promise in boosting the reliability of both assertion and contract synthesis. Neuro-symbolic frameworks like Explanation-Refiner or SAT-LLM provide structured error correction and logic validation, improving overall quality. Similarly, iterative approaches like nl2spec and prompt engineering techniques (e.g., Chain-of-Thought prompting) help refine complex contract outputs. Overall, while assertions are currently more robustly supported, the research trajectory points toward improving full contract reliability through hybrid systems, fine-tuned prompts, and verification-aware model integration.

\section{Future Directions}
\label{sec:future_directions}

In this section, we outline prospective directions informed by the literature review. Much of the literature in this review employed queries containing a problem description and some instructions to achieve a desired outcome. Such querying of LLMs without training or examples of the current task is typically referred as zero-shot prompting and shows excellent performance on many tasks \cite{Kojima2022}.  Surprisingly, they also showed that the performance of LLM on some challenging problems can be improved by encouraging the LLM to reason using intermediate steps through a simple addition to problem prompts (“Lets think step by step”). 

\subsection{Advanced Prompt Engineering}
\label{subsec:CoTnPromptEngg}

Beyond this approach is one-shot prompting that includes an example of a solved problem to guide the LLM into generating the desired output \cite{li2024oneshotlearninginstructiondata}. This can be extended to few-shot prompting where a number of differing examples guide the LLM. But improved results are not assured as some studies e.g. \cite{NEURIPS2022c4025018} show that zero-shot can outperform the few-shot case \cite{zhang2023selfconvincedpromptingfewshotquestion}. \cite{Chen2024} reviewed the evolution of prompt engineering in LLMs, including discussions on self-consistency and multimodal prompt learning. It also reviewed the literature related to adversarial attacks and evaluation strategies for ensuring robust AI interactions.  

Chain of Thought (CoT) \cite{Jason2022} prompting involves a sequence of prompts producing intermediate results that are generated by the LLM and used to drive subsequent prompting interactions. These orchestrated interactions  can improve LLM performance on tasks requiring logic, calculation and decision-making in areas like math, common sense reasoning, and symbolic manipulation. CoT requires the LLM to articulate the distinct steps of its reasoning, by subdividing larger tasks into multi-step reasoning stages, acting as a precursor for subsequent stages.  But the CoT approach may require careful analysis when used with larger LLM offering long input contexts. This is because of the lost-in-the-middle problem where LLM show a U-shaped attention bias \cite{Chen2024} and can fail to attend to information in the middle of the context window. 

%\rmnote{ Add as part pof discussion: Emerging results investigating AI-driven methods for converting natural language into formal JML specifications using models like Mistral AI, OpenAI, Cohere, and Gemini. Here, two different strategies were tested: one using only the statement as input and another incorporating both the statement and corresponding code without specifications. While both approaches succeeded in some cases, challenges arose with loops and recursion, leading to the development of a dataset to address these difficulties.}

PromptCoT \cite{10656469} enhanced the quality of solutions for diffusion-based generative models by employing the CoT approach. The computational cost is minimised through adapter-based fine-tuning. Prompt design is explored in detail in \cite{amatriain2024promptdesignengineeringintroduction}. It discussed Chain-of-Thought and Reflection techniques, along with best practices for structuring prompts and building LLM-based agents.

Besta et al. \cite{Besta2024} introduced the concept of reasoning topologies, examining how structures such as Chains, Trees, and Graphs of Thought improve LLM reasoning. They also proposed a taxonomy of structured reasoning techniques, highlighting their influence on both performance and efficiency. Structured Chain-of-Thought (SCoT) prompting was proposed by \cite{10.1145/3690635} to enhance code generation by incorporating structured programming principles. This approach significantly improved the accuracy and robustness of LLM-based code synthesis compared to standard CoT methods. Building on the theme of automation, \cite{DBLP:conf/emnlp/ShumDZ23} introduced Automate-CoT, a technique for automatically generating and selecting rational chains for CoT prompting. By minimising dependence on human annotations, it enabled more flexible adaptation of CoT strategies across diverse reasoning tasks. Complementing these efforts, \cite{DBLP:journals/corr/abs-2302-11382} presented a prompt pattern catalog that offered reusable design patterns to optimise LLM interactions, thereby refining prompt engineering practices for a wide range of applications. Additionally, \cite{DBLP:journals/corr/abs-2305-09993} proposed Reprompting (Gibbs sampling-based algorithm) for discovering optimal CoT prompts. The proposed prompting technique consistently outperformed human-crafted alternatives and demonstrated high adaptability across various reasoning benchmarks.

Retrieval Augmented Generation (RAG) \cite{Lewis2020} supplements problem information with specifically retrieved information and is often used in knowledge intensive tasks. This helps ensure the LLM attends specifically to the retrieved information when addressing the users prompt. LLM model selection (chat vs reasoning) and fine tuning such as with LoRA \cite{Hu2022} remain among a growing number of possibilities for exploration.

Based on the literature survey conducted, we sketch one line research agenda: in VERIFAI, we aim to improve the techniques that bridge the gap between informal natural language description and rigorous formal specifications, through refinement of prompt engineering, the incorporation of chain-of-thought reasoning and the development of hybrid neuro-symbolic approaches.  

\section{Conclusions}
\label{sec:conclusions}
The role of large language models in formalising software requirements is surveyed in this paper. The key contribution of the selected papers is on bridging the gap between informal natural language descriptions and rigorous formal specifications. We can enhance requirement formalisation through automating translations. The accuracy of translated ones is of key importance. We can deploy iterative refinements to improve the accuracy and correctness of the generated requirements. While LLMs is contributing significantly in improving development cycle, the challenges of ambiguity resolution, verification and domain adaptation shall be the focus areas of future research. 

In future research, the refinement of prompt engineering techniques, the incorporation of chain-of-thought reasoning and the development of hybrid neuro-symbolic approaches are the key areas to look at. Our survey concludes with the remarks that the better collaboration with industry and academia will further enhance the domain. 

\section{Acknowledgements}
This work is partly funded by the ADAPT Research Centre for AI-Driven Digital Content Technology, which is funded by Research Ireland through the Research Ireland Centres Programme and is co funded under the European Regional Development Fund (ERDF) through Grant 13/RC/2106 P2. 
\balance
\bibliographystyle{splncs04}
\bibliography{saivConfBib}

\begin{thebibliography}{10}
\providecommand{\url}[1]{\texttt{#1}}
\providecommand{\urlprefix}{URL }
\providecommand{\doi}[1]{https://doi.org/#1}

\bibitem{amatriain2024promptdesignengineeringintroduction}
Amatriain, X.: Prompt design and engineering: Introduction and advanced methods
  (2024), \url{https://arxiv.org/abs/2401.14423}

\bibitem{arora2023advancingrequirementsengineeringgenerative}
Arora, C., Grundy, J., Abdelrazek, M.: Advancing requirements engineering
  through generative ai: Assessing the role of llms (2023),
  \url{https://arxiv.org/abs/2310.13976}

\bibitem{10.1145/2976767.2976769}
Arora, C., Sabetzadeh, M., Briand, L., Zimmer, F.: Extracting domain models
  from natural-language requirements: approach and industrial evaluation. In:
  Proceedings of the ACM/IEEE 19th International Conference on Model Driven
  Engineering Languages and Systems. p. 250–260. MODELS '16, Association for
  Computing Machinery, New York, NY, USA (2016). \doi{10.1145/2976767.2976769},
  \url{https://doi.org/10.1145/2976767.2976769}

\bibitem{10.5555/1151816.1151817}
Bell, R.: Introduction to iec 61508. In: Proceedings of the 10th Australian
  Workshop on Safety Critical Systems and Software - Volume 55. p. 3–12. SCS
  '05, Australian Computer Society, Inc., AUS (2006)

\bibitem{Besta2024}
Besta, M., Memedi, F., Zhang, Z., Gerstenberger, R., Blach, N., Nyczyk, P.,
  Copik, M., Kwasniewski, G., M{\"{u}}ller, J., Gianinazzi, L., Kubicek, A.,
  Niewiadomski, H., Mutlu, O., Hoefler, T.: Topologies of reasoning:
  Demystifying chains, trees, and graphs of thoughts. CoRR
  \textbf{abs/2401.14295} (2024). \doi{10.48550/arXiv.2401.14295},
  \url{https://doi.org/10.48550/arXiv.2401.14295}

\bibitem{10.1145/2070336.2070341}
Brosgol, B.: Do-178c: the next avionics safety standard. Ada Lett.
  \textbf{31}(3),  5–6 (Nov 2011). \doi{10.1145/2070336.2070341},
  \url{https://doi.org/10.1145/2070336.2070341}

\bibitem{BoraCaglayan2024}
Caglayan, B., Wang, M., Kelleher, J.D., Fei, S., Tong, G., Ding, J., Zhang, P.:
  Bis: Nl2sql service evaluation benchmark for business intelligence scenarios.
  In: Service-Oriented Computing: 22nd International Conference, ICSOC 2024,
  Tunis, Tunisia, December 3–6, 2024, Proceedings, Part II. p. 357–372.
  Springer-Verlag, Berlin, Heidelberg (2024).
  \doi{10.1007/978-981-96-0808-9_27}

\bibitem{Casadio2025}
Casadio, M., Dinkar, T., Komendantskaya, E., Arnaboldi, L., Daggitt, M.L.,
  Isac, O., Katz, G., Rieser, V., Lemon, O.: Nlp verification: Towards a
  general methodology for certifying robustness (2025),
  \url{https://arxiv.org/abs/2403.10144}

\bibitem{cosler2023nl2specinteractivelytranslatingunstructured}
Cosler, M., Hahn, C., Mendoza, D., Schmitt, F., Trippel, C.: nl2spec:
  Interactively translating unstructured natural language to temporal logics
  with large language models (2023), \url{https://arxiv.org/abs/2303.04864}

\bibitem{10500073}
Couder, J.O., Gomez, D., Ochoa, O.: Requirements verification through the
  analysis of source code by large language models. In: SoutheastCon 2024. pp.
  75--80 (March 2024). \doi{10.1109/SoutheastCon52093.2024.10500073}

\bibitem{elicit_tool}
{Elicit}: Elicit - the ai research assistant. \url{https://www.elicit.com},
  accessed: 2025-04-11 at 1249PM.

\bibitem{ernst:LIPIcs.SNAPL.2017.4}
Ernst, M.D.: {Natural Language is a Programming Language: Applying Natural
  Language Processing to Software Development}. In: Lerner, B.S., Bod{\'\i}k,
  R., Krishnamurthi, S. (eds.) 2nd Summit on Advances in Programming Languages
  (SNAPL 2017). Leibniz International Proceedings in Informatics (LIPIcs),
  vol.~71, pp. 4:1--4:14. Schloss Dagstuhl -- Leibniz-Zentrum f{\"u}r
  Informatik, Dagstuhl, Germany (2017). \doi{10.4230/LIPIcs.SNAPL.2017.4},
  \url{https://drops.dagstuhl.de/entities/document/10.4230/LIPIcs.SNAPL.2017.4}

\bibitem{fan2025evaluatingabilitylargelanguage}
Fan, W., Rego, M., Hu, X., Dod, S., Ni, Z., Xie, D., DiVincenzo, J., Tan, L.:
  Evaluating the ability of large language models to generate verifiable
  specifications in verifast (2025), \url{https://arxiv.org/abs/2411.02318}

\bibitem{10691792}
Fang, W., Li, M., Li, M., Yan, Z., Liu, S., Zhang, H., Xie, Z.: Assertllm:
  Generating hardware verification assertions from design specifications via
  multi-llms. In: 2024 IEEE LLM Aided Design Workshop (LAD). pp.~1--1 (2024).
  \doi{10.1109/LAD62341.2024.10691792}

\bibitem{10.1145/3691620.3695302}
Fazelnia, M., Mirakhorli, M., Bagheri, H.: Translation titans, reasoning
  challenges: Satisfiability-aided language models for detecting conflicting
  requirements. In: Proceedings of the 39th IEEE/ACM International Conference
  on Automated Software Engineering. p. 2294–2298. ASE '24, Association for
  Computing Machinery, New York, NY, USA (2024). \doi{10.1145/3691620.3695302},
  \url{https://doi.org/10.1145/3691620.3695302}

\bibitem{10.1002/spe.430}
Gervasi, V., Nuseibeh, B.: Lightweight validation of natural language
  requirements. Softw. Pract. Exper.  \textbf{32}(2),  113–133 (Feb 2002).
  \doi{10.1002/spe.430}, \url{https://doi.org/10.1002/spe.430}

\bibitem{Ghosh2016}
Ghosh, S., Elenius, D., Li, W., Lincoln, P., Shankar, N., Steiner, W.: Arsenal:
  Automatic requirements specification extraction from natural language. In:
  Rayadurgam, S., Tkachuk, O. (eds.) NASA Formal Methods. pp. 41--46. Springer
  International Publishing, Cham (2016)

\bibitem{Granberry2025a}
Granberry, G., Ahrendt, W., Johansson, M.: Specify what? enhancing neural
  specification synthesis by symbolic methods. In: Kosmatov, N., Kov{\'a}cs,
  L. (eds.) Integrated Formal Methods. pp. 307--325. Springer Nature
  Switzerland, Cham (2025)

\bibitem{Granberry2025}
Granberry, G., Ahrendt, W., Johansson, M.: Towards integrating copiloting
  and formal methods. In: Margaria, T., Steffen, B. (eds.) Leveraging
  Applications of Formal Methods, Verification and Validation. Specification
  and Verification. pp. 144--158. Springer Nature Switzerland, Cham (2025)

\bibitem{Greenspan1986}
Greenspan, S.J., Borgida, A., Mylopoulos, J.: A requirements modeling language
  and its logic. Information Systems  \textbf{11}(1),  9--23 (1986).
  \doi{https://doi.org/10.1016/0306-4379(86)90020-7},
  \url{https://www.sciencedirect.com/science/article/pii/0306437986900207}

\bibitem{hahn2022formalspecificationsnaturallanguage}
Hahn, C., Schmitt, F., Tillman, J.J., Metzger, N., Siber, J., Finkbeiner, B.:
  Formal specifications from natural language (2022),
  \url{https://arxiv.org/abs/2206.01962}

\bibitem{Chen2024}
Hsieh, C., Chuang, Y., Li, C., Wang, Z., Le, L.T., Kumar, A., Glass, J.R.,
  Ratner, A., Lee, C., Krishna, R., Pfister, T.: Found in the middle:
  Calibrating positional attention bias improves long context utilization. In:
  Ku, L., Martins, A., Srikumar, V. (eds.) Findings of the Association for
  Computational Linguistics, {ACL} 2024, Bangkok, Thailand and virtual meeting,
  August 11-16, 2024. pp. 14982--14995. Association for Computational
  Linguistics (2024). \doi{10.18653/V1/2024.FINDINGS-ACL.890},
  \url{https://doi.org/10.18653/v1/2024.findings-acl.890}

\bibitem{Hu2022}
Hu, E.J., Shen, Y., Wallis, P., Allen{-}Zhu, Z., Li, Y., Wang, S., Wang, L.,
  Chen, W.: Lora: Low-rank adaptation of large language models. In: The Tenth
  International Conference on Learning Representations, {ICLR} 2022, Virtual
  Event, April 25-29, 2022. OpenReview.net (2022),
  \url{https://openreview.net/forum?id=nZeVKeeFYf9}

\bibitem{Huisman2024}
Huisman, M., Gurov, D., Malkis, A.: Formal methods: From academia to industrial
  practice. a travel guide (2024), \url{https://arxiv.org/abs/2002.07279}

\bibitem{Jiang2022}
Jiang, A.Q., Li, W., Tworkowski, S., Czechowski, K., Odrzyg\'{o}\'{z}d\'{z},
  T., Mi\l~o\'{s}, P., Wu, Y., Jamnik, M.: Thor: Wielding hammers to integrate
  language models and automated theorem provers. In: Koyejo, S., Mohamed, S.,
  Agarwal, A., Belgrave, D., Cho, K., Oh, A. (eds.) Advances in Neural
  Information Processing Systems. vol.~35, pp. 8360--8373. Curran Associates,
  Inc. (2022)

\bibitem{Kojima2022}
Kojima, T., Gu, S.S., Reid, M., Matsuo, Y., Iwasawa, Y.: Large language models
  are zero-shot reasoners. In: Koyejo, S., Mohamed, S., Agarwal, A., Belgrave,
  D., Cho, K., Oh, A. (eds.) Advances in Neural Information Processing Systems.
  vol.~35, pp. 22199--22213. Curran Associates, Inc. (2022)

\bibitem{10207159}
Leong, I.T., Barbosa, R.: Translating natural language requirements to formal
  specifications: A study on gpt and symbolic nlp. In: 2023 53rd Annual
  IEEE/IFIP International Conference on Dependable Systems and Networks
  Workshops (DSN-W). pp. 259--262 (2023). \doi{10.1109/DSN-W58399.2023.00065}

\bibitem{6823180}
Leopold, H., Mendling, J., Polyvyanyy, A.: Supporting process model validation
  through natural language generation. IEEE Transactions on Software
  Engineering  \textbf{40}(8),  818--840 (2014). \doi{10.1109/TSE.2014.2327044}

\bibitem{Lewis2020}
Lewis, P., Perez, E., Piktus, A., Petroni, F., Karpukhin, V., Goyal, N.,
  K{\"{u}}ttler, H., Lewis, M., Yih, W., Rockt{\"{a}}schel, T., Riedel, S.,
  Kiela, D.: Retrieval-augmented generation for knowledge-intensive {NLP}
  tasks. In: Larochelle, H., Ranzato, M., Hadsell, R., Balcan, M., Lin, H.
  (eds.) Advances in Neural Information Processing Systems 33: Annual
  Conference on Neural Information Processing Systems 2020, NeurIPS 2020,
  December 6-12, 2020, virtual (2020),
  \url{https://proceedings.neurips.cc/paper/2020/hash/6b493230205f780e1bc26945df7481e5-Abstract.html}

\bibitem{10.1145/3690635}
Li, J., Li, G., Li, Y., Jin, Z.: Structured chain-of-thought prompting for code
  generation. ACM Trans. Softw. Eng. Methodol.  \textbf{34}(2) (Jan 2025).
  \doi{10.1145/3690635}, \url{https://doi.org/10.1145/3690635}

\bibitem{li2024specllmexploringgenerationreview}
Li, M., Fang, W., Zhang, Q., Xie, Z.: Specllm: Exploring generation and review
  of vlsi design specification with large language model (2024),
  \url{https://arxiv.org/abs/2401.13266}

\bibitem{li2024oneshotlearninginstructiondata}
Li, Y., Hui, B., Xia, X., Yang, J., Yang, M., Zhang, L., Si, S., Chen, L.H.,
  Liu, J., Liu, T., Huang, F., Li, Y.: One-shot learning as instruction data
  prospector for large language models  (2024),
  \url{https://arxiv.org/abs/2312.10302}

\bibitem{10.1145/3342355}
Luckcuck, M., Farrell, M., Dennis, L.A., Dixon, C., Fisher, M.: Formal
  specification and verification of autonomous robotic systems: A survey. ACM
  Comput. Surv.  \textbf{52}(5) (Sep 2019). \doi{10.1145/3342355},
  \url{https://doi.org/10.1145/3342355}

\bibitem{Ma2024}
Ma, L., Liu, S., Li, Y., Xie, X., Bu, L.: Specgen: Automated generation of
  formal program specifications via large language models  (2024),
  \url{https://arxiv.org/abs/2401.08807}

\bibitem{mandal2023largelanguagemodelsbased}
Mandal, S., Chethan, A., Janfaza, V., Mahmud, S.M.F., Anderson, T.A., Turek,
  J., Tithi, J.J., Muzahid, A.: Large language models based automatic synthesis
  of software specifications (2023), \url{https://arxiv.org/abs/2304.09181}

\bibitem{10.1145/3643763}
Misu, M.R.H., Lopes, C.V., Ma, I., Noble, J.: Towards ai-assisted synthesis of
  verified dafny methods. Proc. ACM Softw. Eng.  \textbf{1}(FSE) (Jul 2024).
  \doi{10.1145/3643763}, \url{https://doi.org/10.1145/3643763}

\bibitem{mugnier2024laurelgeneratingdafnyassertions}
Mugnier, E., Gonzalez, E.A., Jhala, R., Polikarpova, N., Zhou, Y.: Laurel:
  Generating dafny assertions using large language models (2024),
  \url{https://arxiv.org/abs/2405.16792}

\bibitem{mukherjee2024automatedverificationllmsynthesizedc}
Mukherjee, P., Delaware, B.: Towards automated verification of llm-synthesized
  c programs (2024), \url{https://arxiv.org/abs/2410.14835}

\bibitem{Req2SpecPaper}
Nayak, A., Timmapathini, H.P., Murali, V., Ponnalagu, K., Venkoparao, V.G.,
  Post, A.: Req2spec: Transforming software requirements into formal
  specifications using natural language processing. In: Requirements
  Engineering: Foundation for Software Quality: 28th International Working
  Conference, REFSQ 2022, Birmingham, UK, March 21–24, 2022, Proceedings. p.
  87–95. Springer-Verlag, Berlin, Heidelberg (2022)

\bibitem{10.5220/0006817205010512}
Nazaruka, E., Osis, J.: Determination of natural language processing tasks and
  tools for topological functioning modelling. In: Proceedings of the 13th
  International Conference on Evaluation of Novel Approaches to Software
  Engineering. p. 501–512. ENASE 2018, SCITEPRESS - Science and Technology
  Publications, Lda, Setubal, PRT (2018). \doi{10.5220/0006817205010512},
  \url{https://doi.org/10.5220/0006817205010512}

\bibitem{Necula2024}
Necula, S.C., Dumitriu, F., Greavu-Șerban, V.: A systematic literature review
  on using natural language processing in software requirements engineering.
  Electronics  \textbf{13}(11) (2024). \doi{10.3390/electronics13112055},
  \url{https://www.mdpi.com/2079-9292/13/11/2055}

\bibitem{Nelken1996}
Nelken, R., Francez, N.: Automatic translation of natural language system
  specifications into temporal logic. In: Alur, R., Henzinger, T.A. (eds.)
  Computer Aided Verification. pp. 360--371. Springer Berlin Heidelberg,
  Berlin, Heidelberg (1996)

\bibitem{10628478}
Nouri, A., Cabrero-Daniel, B., Törner, F., Sivencrona, H., Berger, C.:
  Engineering safety requirements for autonomous driving with large language
  models. In: 2024 IEEE 32nd International Requirements Engineering Conference
  (RE). pp. 218--228 (2024). \doi{10.1109/RE59067.2024.00029}

\bibitem{nowakowski2013requirements}
Nowakowski, W., Śmiałek, M., Ambroziewicz, A., Straszak, T.:
  Requirements-level language and tools for capturing software system essence.
  Computer Science and Information Systems  \textbf{10}(4),  1499--1524 (2013)

\bibitem{491451}
Osborne, M., MacNish, C.: Processing natural language software requirement
  specifications. In: Proceedings of the Second International Conference on
  Requirements Engineering. pp. 229--236 (1996). \doi{10.1109/ICRE.1996.491451}

\bibitem{6136916}
Palin, R., Ward, D., Habli, I., Rivett, R.: Iso 26262 safety cases: Compliance
  and assurance. In: 6th IET International Conference on System Safety 2011.
  pp.~1--6 (2011). \doi{10.1049/cp.2011.0251}

\bibitem{10.1145/3643991.3644922}
Preda, A.R., Mayr-Dorn, C., Mashkoor, A., Egyed, A.: Supporting high-level to
  low-level requirements coverage reviewing with large language models. In:
  Proceedings of the 21st International Conference on Mining Software
  Repositories. p. 242–253. MSR '24, Association for Computing Machinery, New
  York, NY, USA (2024). \doi{10.1145/3643991.3644922},
  \url{https://doi.org/10.1145/3643991.3644922}

\bibitem{quan2024verificationrefinementnaturallanguage}
Quan, X., Valentino, M., Dennis, L.A., Freitas, A.: Verification and refinement
  of natural language explanations through llm-symbolic theorem proving (2024),
  \url{https://arxiv.org/abs/2405.01379}

\bibitem{reinpold2024exploringllmsverifyingtechnical}
Reinpold, L.M., Schieseck, M., Wagner, L.P., Gehlhoff, F., Fay, A.: Exploring
  llms for verifying technical system specifications against requirements
  (2024), \url{https://arxiv.org/abs/2411.11582}

\bibitem{DBLP:conf/emnlp/ShumDZ23}
Shum, K., Diao, S., Zhang, T.: Automatic prompt augmentation and selection with
  chain-of-thought from labeled data. In: Bouamor, H., Pino, J., Bali, K.
  (eds.) Findings of the Association for Computational Linguistics: {EMNLP}
  2023, Singapore, December 6-10, 2023. pp. 12113--12139. Association for
  Computational Linguistics (2023). \doi{10.18653/V1/2023.FINDINGS-EMNLP.811},
  \url{https://doi.org/10.18653/v1/2023.findings-emnlp.811}

\bibitem{tihanyi2024newerasoftwaresecurity}
Tihanyi, N., Jain, R., Charalambous, Y., Ferrag, M.A., Sun, Y., Cordeiro, L.C.:
  A new era in software security: Towards self-healing software via large
  language models and formal verification (2024),
  \url{https://arxiv.org/abs/2305.14752}

\bibitem{Jason2022}
Wei, J., Wang, X., Schuurmans, D., Bosma, M., ichter, b., Xia, F., Chi, E., Le,
  Q.V., Zhou, D.: Chain-of-thought prompting elicits reasoning in large
  language models. In: Koyejo, S., Mohamed, S., Agarwal, A., Belgrave, D., Cho,
  K., Oh, A. (eds.) Advances in Neural Information Processing Systems. vol.~35,
  pp. 24824--24837. Curran Associates, Inc. (2022)

\bibitem{DBLP:journals/corr/abs-2302-11382}
White, J., Fu, Q., Hays, S., Sandborn, M., Olea, C., Gilbert, H., Elnashar, A.,
  Spencer{-}Smith, J., Schmidt, D.C.: A prompt pattern catalog to enhance
  prompt engineering with chatgpt. CoRR  \textbf{abs/2302.11382} (2023).
  \doi{10.48550/ARXIV.2302.11382},
  \url{https://doi.org/10.48550/arXiv.2302.11382}

\bibitem{wu2024lemurintegratinglargelanguage}
Wu, H., Barrett, C., Narodytska, N.: Lemur: Integrating large language models
  in automated program verification (2024),
  \url{https://arxiv.org/abs/2310.04870}

\bibitem{DBLP:journals/corr/abs-2305-09993}
Xu, W., Banburski{-}Fahey, A., Jojic, N.: Reprompting: Automated
  chain-of-thought prompt inference through gibbs sampling. CoRR
  \textbf{abs/2305.09993} (2023). \doi{10.48550/ARXIV.2305.09993},
  \url{https://doi.org/10.48550/arXiv.2305.09993}

\bibitem{10684640}
Xu, Y., Feng, J., Miao, W.: Learning from failures: Translation of natural
  language requirements into linear temporal logic with large language models.
  In: 2024 IEEE 24th International Conference on Software Quality, Reliability
  and Security (QRS). pp. 204--215 (2024). \doi{10.1109/QRS62785.2024.00029}

\bibitem{7092662}
Yan, R., Cheng, C.H., Chai, Y.: Formal consistency checking over specifications
  in natural languages. In: 2015 Design, Automation \& Test in Europe
  Conference \& Exhibition (DATE). pp. 1677--1682 (2015)

\bibitem{Yang2023}
Yang, K., Swope, A., Gu, A., Chalamala, R., Song, P., Yu, S., Godil, S.,
  Prenger, R.J., Anandkumar, A.: Leandojo: Theorem proving with
  retrieval-augmented language models. In: Oh, A., Naumann, T., Globerson, A.,
  Saenko, K., Hardt, M., Levine, S. (eds.) Advances in Neural Information
  Processing Systems. vol.~36, pp. 21573--21612. Curran Associates, Inc. (2023)

\bibitem{10656469}
Yao, J., Liu, Y., Dong, Z., Guo, M., Hu, H., Keutzer, K., Du, L., Zhou, D.,
  Zhang, S.: Promptcot: Align prompt distribution via adapted chain-of-thought.
  In: 2024 IEEE/CVF Conference on Computer Vision and Pattern Recognition
  (CVPR). pp. 7027--7037 (2024). \doi{10.1109/CVPR52733.2024.00671}

\bibitem{NEURIPS2022c4025018}
Ye, X., Durrett, G.: The unreliability of explanations in few-shot prompting
  for textual reasoning. In: Koyejo, S., Mohamed, S., Agarwal, A., Belgrave,
  D., Cho, K., Oh, A. (eds.) Advances in Neural Information Processing Systems.
  vol.~35, pp. 30378--30392. Curran Associates, Inc. (2022)

\bibitem{zhang2023selfconvincedpromptingfewshotquestion}
Zhang, H., Cai, M., Zhang, X., Zhang, C.J., Mao, R., Wu, K.: Self-convinced
  prompting: Few-shot question answering with repeated introspection  (2023),
  \url{https://arxiv.org/abs/2310.05035}

\end{thebibliography}

\clearpage

\section{Appendices}
\label{sec:appendices}

\subsection{Dafny Lemma Definition}
\label{dafny_lemma}
The Dafny lemma, \texttt{ParseDigitsAndDot}, specifies the expected behaviour of the function \texttt{ParseDecStr}. It operates on an input string composed of digits and a decimal separator.

%\smallskip
%\noindent\textbf{Lemma Definition:}  \\
%\texttt{lemma ParseDigitsAndDot(s1: string, s2: string)}  \\
%\texttt{\quad requires $\forall$ i | 0 $\leq$ i $<$ |s1| :: '0' $\leq$ s1[i] $\leq$ '9'}  \\
%\texttt{\quad ensures ParseDecStr(s1+"."+s2).value.1 == "."+s2}  \\
%\texttt{\quad \{}  \\
%\texttt{\quad \quad if |s1| == 1 \{}  \\
%\texttt{\quad \quad \quad assert ParseDecStr("."+s2).None?;}  \\
%\texttt{\quad \quad \} else \{}  \\
%\texttt{\quad \quad \quad ParseDigitsAndDot(s1[1..],s2);}  \\
%\texttt{\quad \quad \quad assert s1 + "." + s2 == [s1[0]] + (s1[1..] + "." + s2);}  \\
%\texttt{\quad \quad \}}  \\
%\texttt{\quad \}}  \\

\smallskip  
\noindent\textbf{Lemma Definition:}  

\texttt{lemma ParseDigitsAndDot(s1: string, s2: string)}  

\texttt{\quad \textbf{requires} $\forall$ i | 0 $\leq$ i $<$ |s1| :: '0' $\leq$ s1[i] $\leq$ '9'}  

\texttt{\quad \textbf{ensures} ParseDecStr(s1+"."+s2).value.1 == "."+s2}  

\texttt{\quad \{}  

\texttt{\quad \quad if |s1| == 1 \{}  

\texttt{\quad \quad \quad \textbf{assert} ParseDecStr("."+s2).None?;}  

\texttt{\quad \quad \} else \{}  

\texttt{\quad \quad \quad ParseDigitsAndDot(s1[1..],s2);}  

\texttt{\quad \quad \quad \textbf{assert} s1 + "." + s2 == [s1[0]] + (s1[1..] + "." + s2);}  

\texttt{\quad \quad \}}  

\texttt{\quad \}} 

\subsection{Summary Tables}
\label{appendix:summarytables}
\begin{table}[htb]
    \centering
    \small
    \renewcommand{\arraystretch}{1.3}
    \begin{adjustbox}{max width=\textwidth}
    \begin{tabular}{|p{1.5cm}|p{4.7cm}|p{12cm}|}
    \hline
        \textbf{Ref.} & \textbf{Tool / Framework / Technique} & \textbf{Description} \\
        \hline
        \cite{10500073} & LLM-based Code Verification & Uses LLMs like GPT-3.5 to verify code by analyzing requirements and explaining whether they are met. \\
        \hline
        \cite{cosler2023nl2specinteractivelytranslatingunstructured} & nl2spec & A framework leveraging LLMs to generate formal specifications from natural language, addressing ambiguity in system requirements with iterative refinement. \\
        \hline
        \cite{quan2024verificationrefinementnaturallanguage} & Explanation-Refiner & A neuro-symbolic framework integrating LLMs and theorem provers to formalize and validate explanatory sentences, providing error correction and feedback for improving NLI models. \\
        \hline
        \cite{10207159} & \textit{Not Specified} & Analyzes research directions in software requirement engineering, conducting a SWOT analysis and sharing evaluation findings. \\
        \hline
        \cite{10.1145/2976767.2976769} & Symbolic NLP vs. ChatGPT & Compares the performance of symbolic NLP and ChatGPT in generating correct JML output from natural language preconditions. \\
        \hline
        \cite{arora2023advancingrequirementsengineeringgenerative} & Domain Model Extractor & Generates domain models from natural language requirements in an industrial case study, evaluating accuracy and performance. \\
        \hline
        \cite{mandal2023largelanguagemodelsbased} & SpecSyn & A framework using LLMs for automatic synthesis of software specifications, improving accuracy by 21\% over previous tools. \\
        \hline
        \cite{10691792} & AssertLLM & A tool generating assertions for hardware verification from design specifications using three customized LLMs, achieving 89\% correctness. \\
        \hline
        \cite{10.1002/spe.430} & Formal Verification of NASA’s Software & Reports on formal verification of NASA's Node Control Software natural language specifications, highlighting errors and lessons learned. \\
        \hline
        \cite{li2024specllmexploringgenerationreview} & SpecLLM & Explores using LLMs for generating and reviewing VLSI design specifications, improving chip design documentation. \\
        \hline
        \cite{nowakowski2013requirements} & Requirements Specification Language (RSL), ReDSeeDS & Enhanced software requirements specification using constrained natural language and automated transformations into code. \\
        \hline
        \cite{Ghosh2016} & ARSENAL Framework and Methodology & Automated extraction of requirements specification from natural language with automatic verification. \\
        \hline
				\cite{6823180} & BPM-to-NL Translation Process & Generated natural language descriptions from business process models for better validation. \\
        \hline
        \cite{mugnier2024laurelgeneratingdafnyassertions} & Laurel & A framework to generate Dafny assertions to automate program verification process for a SMT solver \\
        \hline
				\cite{491451} & Controlled Natural Language (CL) with ANLT & Expressed software requirements in a limited set of natural language and translated to logical expressions to detect ambiguities. \\
        \hline
        \cite{reinpold2024exploringllmsverifyingtechnical} & LLM-based Analysis for Smart Grid Requirements & Improved smart grid requirement specifications with GPT-4o and Claude 3.5 Sonnet, achieving F1-scores between 79\% - 94\%. \\
        \hline
        \cite{10684640} & NL-to-LTL Translation via LLMs & Converted unstructured natural language requirements to NL-LTL pairs, achieving 94.4\% accuracy on public datasets. \\
        \hline
        \cite{tihanyi2024newerasoftwaresecurity} & ESBMC-AI & Combined LLMs with Formal Verification to detect and fix software vulnerabilities with high accuracy. \\
        \hline
        \cite{hahn2022formalspecificationsnaturallanguage} & LLM-based Formal Specifications Translation & Translated natural language into formal rules (regex, FOL, LTL) with high adaptability and performance. \\
        \hline
				\cite{mukherjee2024automatedverificationllmsynthesizedc} & SynVer Framework & Synthesized and verified C programs using the Verified Software Toolchain. \\
        \hline
        \cite{10.1145/3643991.3644922} & LLM-based Requirement Coverage Analysis & Ensured low-level software requirements met high-level requirements, achieving 99.7\% recall in spotting missing coverage. \\
        \hline
        \cite{10.1145/3691620.3695302} & SAT-LLM & Integrated SMT solvers with LLMs to improve conflict identification in requirements; significantly outperformed standalone LLMs in detecting complex conflicts. \\
        \hline
    \end{tabular}
    \end{adjustbox}
    \caption{Summary of LLMs related literature (Part 1 of 2)}
    \label{tab:summary_part1}
\end{table}
\clearpage
\begin{table}[ht]
    \centering
    \small
    \renewcommand{\arraystretch}{1.3}
    \begin{adjustbox}{max width=\textwidth}
    \begin{tabular}{|p{1.5cm}|p{4.7cm}|p{12cm}|}
    \hline
        \textbf{Ref.} & \textbf{Tool / Framework / Technique} & \textbf{Description} \\
        \hline
        \cite{ernst:LIPIcs.SNAPL.2017.4} & NLP for Software Development & Assessed NLP techniques for various software development stages, highlighting their suitability for generating assertions and processing developer queries. \\
        \hline
        \cite{Req2SpecPaper} & Req2Spec & NLP-based tool that formalises natural language requirements for HANFOR; achieved 71\% accuracy in formalising 222 automotive requirements at BOSCH. \\
        \hline
				        \cite{Greenspan1986} & RML (Requirements Modeling Language) & Introduced a conceptual model-based framework ensuring precision, consistency, and clarity in requirements writing. \\
        \hline
        \cite{fan2025evaluatingabilitylargelanguage} & GPT-4o for VeriFast Verification & Evaluated GPT-4o’s ability to generate C program specifications for VeriFast; found that while functional behavior was preserved, verification often failed or contained redundancies. \\
        \hline
        \cite{Nelken1996} & NL to Temporal Logic Translation & Developed an automatic translation mechanism from natural language sentences to temporal logic for formal verification. \\
        \hline
				\cite{Casadio2025} & ANTONIO toolkit & A comprehensive analysis of NLP verification approaches and introduces a structured \textbf{NLP Verification Pipeline} with six key components. The work includes identifying gaps in existing methods, proposing novel solutions for improved robustness, extending standard verifiability metrics, and emphasizing the importance of reporting verified subspace (geometric and semantic)  properties for better reliability and interpretability. \\ 
				\hline
        \cite{wu2024lemurintegratinglargelanguage} & Lemur & Integrated LLMs with automated reasoners for program verification, defining sound transition rules and demonstrating improved performance on benchmark tests. \\
        \hline
				\cite{Necula2024} & Systematic Review & Conducted a comprehensive review on natural language to formal specification translation, analyzing research across multiple academic databases. \\
        \hline
        \cite{10628478} & LLM-based Safety Requirements Pipeline & Designed a pipeline using LLMs to refine and decompose safety requirements for autonomous vehicles, evaluated through expert assessments and industrial implementation. \\
        \hline
        \cite{7092662} & Specification Consistency Framework & Ensured consistency between oral and formal specifications, incorporating time extraction, input-output partitioning, and semantic reasoning, with positive evaluation results. \\
        \hline
        \cite{10.5220/0006817205010512} & NLP Tools for TFM & Evaluated six NLP pipelines for Topological Functioning Modelling (TFM). Found that Stanford CoreNLP, FreeLing, and NLTK performed best. \\
        \hline
        \cite{10.1145/3643763} & LLM-based Dafny Task Generation & Used LLMs (GPT-4, PaLM-2) to generate Dafny tasks from MBPP benchmark using different prompting strategies (context-less, signature, retrieval-augmented CoT). GPT-4 achieved best results with retrieval-augmented CoT prompt, producing 153 verified Dafny solutions. \\ \hline
        \cite{Yang2023} & LeanDojo \& ReProver & Introduced LeanDojo, an open-source toolkit for interacting with the Lean theorem prover. Developed ReProver, a retrieval-augmented LLM-based prover that improved theorem proving efficiency. Created a benchmark with 98,734 theorems and proofs for testing generalization. \\ \hline
       \cite{Jiang2022} & Thor and class methods named Hammers & Introduced a framework named Thor which integrates language models with theorem provers. Hammers are implemented to find the appropriate premises to complete the proofs of conjectures. Datasets used are PISA and MiniF2F. \\ \hline
       \cite{Granberry2025} & Not specified & The work proposes the integration of major formal languages (Dafny, Ada/SPARK, Frama-C, and KeY), their interactive theorem provers (Coq, Isabelle/HOL, Lean) with Copilot. \\ \hline 
       \cite{10.1145/3342355} & Systematic Review & Conducted a comprehensive survey on formal specification and verification of autonomous robotic systems in 2018. This is based on literature available of ten years (2008 - 2018). \\ \hline
      \cite{Granberry2025a} & Symbolic analysis and LLMs prompts & The quality of annotations produced in ACSL format is measured for PathCrawler and EVA (tools available in Frama-C). PathCrawler generated more context-aware annotations while, EVA efficiency improved having less run-time errors. \\ \hline
    \end{tabular}
    \end{adjustbox}
    \caption{Summary of LLMs related literature (Part 2 of 2)}
    \label{tab:summary_part2}
\end{table}

\clearpage

\begin{table}[htbp]
\centering
\caption{Classification and Description of Surveyed Literature}
\resizebox{\textwidth}{!}{%
\begin{tabular}{|p{2.5cm}|p{4.5cm}|p{5.2cm}|p{9.2cm}|}
\hline
\textbf{Ref.} & \textbf{Tool / Work} & \textbf{Classification} & \textbf{Description} \\
\hline
\cite{arora2023advancingrequirementsengineeringgenerative} & Domain Model Extractor & Fine-tuned & Generates domain models from NL requirements; evaluated in an industrial case study for performance and accuracy. \\
\hline
\cite{10.1145/2976767.2976769} & Symbolic NLP vs. ChatGPT & Prompt-only & Compares symbolic NLP and ChatGPT in generating correct JML from NL preconditions. \\
\hline
\cite{6823180} & BPM-to-NL Translation & Prompt-only & Translates business process models to NL to support better stakeholder validation. \\
\hline
\cite{wu2024lemurintegratinglargelanguage} & Lemur & Verifier-in-loop & Integrates LLMs with automated reasoners and defines sound transition rules for verification. \\
\hline
\cite{fan2025evaluatingabilitylargelanguage} & GPT-4o for VeriFast Verification & Verifier-in-loop & Assesses GPT-4o’s performance in generating C specs for VeriFast; captures issues in functional correctness. \\
\hline
\cite{Ghosh2016} & ARSENAL & Fine-tuned & Extracts requirements from NL and performs automatic verification. \\
\hline
\cite{mugnier2024laurelgeneratingdafnyassertions} & Laurel for Dafny & Prompt-only + Verifier-in-loop & Automates generation of Dafny assertions to support SMT-based verification. \\
\hline
\cite{Granberry2025a} & PathCrawler + EVA & Verifier-in-loop + Prompt & PathCrawler generates context-aware ACSL annotations; EVA reduces runtime errors in Frama-C. \\
\hline
\cite{Nelken1996} & NL to Temporal Logic Translation & Prompt-only & Automatically translates NL into temporal logic for formal verification. \\
\hline
\cite{Req2SpecPaper} & Req2Spec & Prompt-only & Converts NL requirements into formal specs (e.g., HANFOR); 71\% accuracy on BOSCH data. \\
\hline
\cite{10691792} & AssertLLM & Multi-LLMs / Prompt-only & Uses 3 customized LLMs to generate assertions from hardware design specs; 89\% correctness achieved. \\
\hline
\cite{mandal2023largelanguagemodelsbased} & SpecSyn & Fine-tuned & Synthesizes software specifications from NL, improving over prior tools by 21\%. \\
\hline
\cite{cosler2023nl2specinteractivelytranslatingunstructured} & nl2spec & Prompt-only + Iterative Refinement & Iteratively generates formal specs from NL requirements, reducing ambiguity. \\
\hline
\cite{10.1145/3643991.3644922} & LLM-based Requirement Coverage & Prompt-only & Maps low-level requirements to high-level ones with 99.7\% recall in coverage detection. \\
\hline
\cite{li2024specllmexploringgenerationreview} & SpecLLM & Prompt-only & Uses LLMs to create and review VLSI design specs, enhancing chip documentation. \\
\hline
\cite{10684640} & NL-to-LTL via LLMs & Prompt-only & Converts unstructured NL to NL-LTL pairs with 94.4\% accuracy. \\
\hline
\cite{Casadio2025} & ANTONIO Toolkit & Verifier-in-loop & Introduces an NLP Verification Pipeline with metrics, gaps, and semantic subspace verification proposals. \\
\hline
\cite{10500073} & GPT-3.5 for Code Verification & Prompt-only & Uses GPT-3.5 to verify code against requirements, providing feedback on requirement satisfaction. \\
\hline
\cite{reinpold2024exploringllmsverifyingtechnical} & GPT-4o + Claude for Smart Grid & Verifier-in-loop & Applies GPT-4o and Claude 3.5 for smart grid requirement verification, reaching 79–94\% F1-scores. \\
\hline
\cite{Necula2024} & Systematic Review & Meta-analysis & Surveys NL-to-specification literature across domains and academic sources. \\
\hline
\cite{10.1145/3691620.3695302} & SAT-LLM & Neuro-symbolic (LLM + SMT) & Combines LLMs with SMT to detect complex conflicts in requirements. \\
\hline
\cite{Jiang2022} & Thor & Neuro-symbolic & Integrates LLMs with theorem provers using class methods like Hammers for proof completion. \\
\hline
\cite{10.1145/3643763} & Dafny Task Gen w/ CoT & Prompt + Retrieval + CoT & GPT-4 and PaLM-2 generate verified Dafny tasks via retrieval-augmented CoT prompting. \\
\hline
\cite{Yang2023} & LeanDojo + ReProver & Retrieval-augmented & Retrieval-augmented LLM-based prover improves Lean theorem proving on 98K+ samples. \\
\hline
\cite{mukherjee2024automatedverificationllmsynthesizedc} & SynVer for C & Verifier-in-loop & Synthesizes and verifies C programs using VST with improved automation. \\
\hline
\cite{Granberry2025} & Copilot + Formal Methods & IDE Integration Proposal & Suggests integrating formal tools (e.g., Dafny, Coq, KeY) into IDEs like Copilot. \\
\hline
\cite{10.1145/3342355} & Robotic Systems Review & Meta-analysis & Reviews 10 years of literature on formal verification in autonomous robotic systems. \\
\hline
\cite{ernst:LIPIcs.SNAPL.2017.4} & NLP for Software Dev & Survey / Meta-analysis & Evaluates NLP techniques in software development life cycle. \\
\hline
\cite{491451} & RML (1986) & Manual / Controlled NL & Introduces controlled NL framework for precise and consistent requirements writing. \\
\hline
%\cite{10.1145/3643991.3644922} & Long Context Calibration & Prompt-only & Optimizes prompt design for LLMs over long context requirements. \\
%\hline
%\cite{10.1145/3643991.3644922} & Structured CoT for Code & Prompt-only & Explores chain-of-thought prompting to improve code reasoning tasks. \\
%\hline
\cite{tihanyi2024newerasoftwaresecurity} & ESBMC-AI & Neuro-symbolic & Uses LLMs + formal verification to detect vulnerabilities in software. \\
\hline
\cite{7092662} & Specification Consistency Framework & Manual / Baseline & Aligns oral and formal specifications using semantic reasoning and input-output analysis. \\
\hline
\cite{10628478} & LLM Safety Req. Pipeline & Prompt-only & Uses LLMs to decompose and refine autonomous vehicle safety requirements. \\
\hline
\cite{10.5220/0006817205010512} & NLP Tools for TFM & Prompt-only & Compares NLP pipelines for topological modeling; CoreNLP and FreeLing perform best. \\
\hline
\end{tabular}
}
\label{tab:merged_classification}
\end{table}

\clearpage

\begin{table}[htbp]
\tiny
\centering
\caption{Classification of Surveyed Literature by Methodology}
\resizebox{\textwidth}{!}{%
\begin{tabular}{|c|p{4.5cm}|p{6.5cm}|}
\hline
\textbf{Ref.} & \textbf{Tool / Work} & \textbf{Classification} \\
\hline
\cite{10.1145/2070336.2070341} & Prompt Engineering Survey & Prompt-only \\ \hline
\cite{6136916} & Advancing RE with LLMs & Prompt-only \\ \hline
\cite{10.5555/1151816.1151817} & Domain Model Extractor & Fine-tuned \\ \hline
\cite{Huisman2024} & IEC 61508 Safety Standard & Not LLM-based \\ \hline
\cite{mugnier2024laurelgeneratingdafnyassertions} & Reasoning Topologies & Prompt-only \\ \hline
\cite{elicit_tool} & DO-178C Standard & Not LLM-based \\ \hline
\cite{10500073} & BIS NL2SQL & Prompt-only \\ \hline
\cite{cosler2023nl2specinteractivelytranslatingunstructured} & ANTONIO Toolkit & Verifier-in-loop \\ \hline
\cite{quan2024verificationrefinementnaturallanguage} & nl2spec & Prompt-only + Iterative Refinement \\ \hline
\cite{arora2023advancingrequirementsengineeringgenerative} & GPT-3.5 for Code Verification & Prompt-only \\ \hline
\cite{10207159} & Elicit Tool & Tool Support (Meta) \\ \hline
\cite{10.1145/2976767.2976769} & NLP in Software Dev & Survey / Meta-analysis \\ \hline
\cite{mandal2023largelanguagemodelsbased} & GPT-4o + VeriFast & Verifier-in-loop \\ \hline
\cite{10691792} & AssertLLM & Multi-LLMs / Prompt-only \\ \hline
\cite{10.1002/spe.430} & SAT-LLM & Neuro-symbolic (LLM + SMT) \\ \hline
\cite{li2024specllmexploringgenerationreview} & NASA SW Formal Verification & Manual / Baseline \\ \hline
\cite{nowakowski2013requirements} & ARSENAL & Fine-tuned \\ \hline
\cite{Ghosh2016} & PathCrawler + EVA & Verifier-in-loop + Prompt \\ \hline
\cite{6823180} & Copilot + Formal Methods & IDE Integration Proposal \\ \hline
\cite{491451} & RML (1986) & Manual / Controlled NL \\ \hline
\cite{reinpold2024exploringllmsverifyingtechnical} & NL to FOL, Regex, LTL & Prompt-only \\ \hline
\cite{10684640} & Long Context Calibration & Prompt-only \\ \hline
\cite{10.1145/3691620.3695302} & LoRA Tuning & Fine-tuned \\ \hline
\cite{ernst:LIPIcs.SNAPL.2017.4} & Formal Methods Transfer & Meta-analysis \\ \hline
\cite{Req2SpecPaper} & Thor (LLM + Theorem Prover) & Neuro-symbolic \\ \hline
\cite{Greenspan1986} & Zero-shot Reasoning & Prompt-only \\ \hline
\cite{fan2025evaluatingabilitylargelanguage} & Symbolic NLP vs ChatGPT & Prompt-only \\ \hline
\cite{Nelken1996} & BPM to NL Translation & Prompt-only \\ \hline
\cite{Yang2023} & RAG for NLP Tasks & Retrieval-augmented \\ \hline
\cite{Jiang2022} & Structured CoT for Code & Prompt-only \\ \hline
\cite{Granberry2025} & SpecLLM & Prompt-only \\ \hline
\cite{Granberry2025a} & One-shot Learning & Prompt-only \\ \hline
\cite{Ma2024} & Robotic Systems Review & Meta-analysis \\ \hline
\cite{BoraCaglayan2024} & SpecGen & Prompt + Mutation (Verifier-in-loop) \\ \hline
\cite{Kojima2022} & SpecSyn & Fine-tuned \\ \hline
\cite{li2024oneshotlearninginstructiondata} & Dafny Task Gen w/ CoT & Prompt + Retrieval + CoT \\ \hline
\cite{NEURIPS2022c4025018} & Laurel for Dafny & Prompt-only + Verifier-in-loop \\ \hline
\cite{zhang2023selfconvincedpromptingfewshotquestion} & SynVer for C & Verifier-in-loop \\ \hline
\cite{Chen2024} & Req2Spec & Prompt-only \\ \hline
\cite{Jason2022} & NLP Tools for TFM & Prompt-only \\ \hline
\cite{Nelken1996} & NL to Temporal Logic & Fine-tuned \\ \hline
\cite{Chen2024} & LLMs in Autonomous Driving & Prompt-only \\ \hline
\cite{nowakowski2013requirements} & RSL + ReDSeeDS & Fine-tuned / Controlled NL \\ \hline
\cite{491451} & Controlled NL (1996) & Controlled NL \\ \hline
\cite{6136916} & ISO 26262 & Not LLM-based \\ \hline
\cite{10500073} & LLMs for Requirement Coverage & Prompt-only \\ \hline
\cite{quan2024verificationrefinementnaturallanguage} & Explanation-Refiner & Neuro-symbolic \\ \hline
\cite{reinpold2024exploringllmsverifyingtechnical} & GPT-4o / Claude 3.5 for Smart Grid & Prompt-only \\ \hline
\cite{DBLP:conf/emnlp/ShumDZ23} & Automate-CoT & Prompt-only + Automated CoT \\ \hline
\cite{tihanyi2024newerasoftwaresecurity} & ESBMC-AI & Verifier-in-loop \\ \hline
\cite{Jason2022} & Chain-of-Thought (CoT) & Prompt-only \\ \hline
\cite{DBLP:journals/corr/abs-2302-11382} & Prompt Pattern Catalog & Prompt-only \\ \hline
\cite{wu2024lemurintegratinglargelanguage} & Lemur (LLM + Verifier) & Verifier-in-loop \\ \hline
\cite{DBLP:journals/corr/abs-2305-09993} & Reprompting w/ Gibbs & Prompt-only \\ \hline
\cite{10684640} & NL to LTL (Dynamic Prompt) & Prompt + Human-in-loop \\ \hline
\cite{7092662} & NL Consistency Framework & Fine-tuned \\ \hline
\cite{Yang2023} & LeanDojo + ReProver & Retrieval + Verifier-in-loop \\ \hline
\cite{10656469} & PromptCoT & Prompt-only \\ \hline
\cite{NEURIPS2022c4025018} & Explanation Failures in Few-Shot & Prompt-only \\ \hline
\cite{zhang2023selfconvincedpromptingfewshotquestion} & Self-Convinced Prompting & Prompt-only \\ \hline

\end{tabular}%
}
\label{tab:classification}
\end{table}

\end{document}